# Open questions regarding the arrow of time      Sep09

H. D. Zeh – www.zeh-hd.de

Universität Heidelberg, Germany

**Abstract:** Conceptual problems regarding the arrow of time in classical physics, quantum physics, cosmology, and quantum gravity are discussed. Particular attention is paid to the dynamical rôle of the quantum indeterminism, and to various concepts of timelessness.

### 1. Laws and facts

The Second Law of Thermodynamics is usually regarded as the major physical manifestation of the arrow of time, from which many other consequences can be derived. I have discussed the relations between these different forms of the arrow in detail elsewhere,[1] so I will occasionally refer to this source in the following by TD ("Time Direction") for short. This article is meant to review some open conceptual problems, which are often insufficiently realized or otherwise of actual interest.

    In Statistical Thermodynamics, the Second Law is derived from the assumption that a closed system must evolve towards a more probable state. In this context, entropy is defined as a measure of probability. This explanation is incomplete for various reasons. First, the concept of evolution already presumes a direction in time. To regard it as a direction *of* time would even apply this presumption to the very definition of time. This would go beyond a purely mechanistic concept of time, which is defined in accordance with time-symmetric laws of motion. Newton's absolute "flow of time" is a metaphor; its direction would be physically meaningful only if one assumed asymmetric laws. For example, Newton regarded friction as representing a fundamental force that would slow down all motion. So in his opinion God had to intervene once in a while to set things in motion again. Without such an external, meta-physical, or at least law-like fundamental direction in or of time, one can only speak of an asymmetry of the facts (which must and can be compatible with symmetric laws).

    Second, the concept of probabilities requires a measure that is usually defined with respect to ensembles of *possible* states. Since in classical physics every system is assumed at any time to be in one definite microscopic state, the latter must be "coarse-grained" in order to

define a "macroscopic" or "thermodynamical state" (that is, an ensemble of microscopic states) that may then possess a non-trivial probability measure. Various kinds of coarse-graining (omissions of actual or possible information) have been discussed in the rich literature on this subject, or were simply invented in the context of a new theory. The justification of such ensembles by a macroscopic (that is, incompletely defined) preparation procedure would refer to the time-directed concept of preparations as a *deus ex machina* (similar to Newton's divine interventions). Similarly, incomplete observability or controllability of certain degrees of freedom would refer to the limited abilities of external time-asymmetric observers.

The mechanistic concept of time is usually postulated together with the deterministic dynamical laws that are assumed on empirical grounds to control the facts. Eugene P. Wigner has called the distinction between laws and initial conditions (initial facts) Newton's greatest discovery. In a deterministic theory, initial conditions could as well be replaced by final ones, or by conditions at any intermediate time. This mechanistic concept requires only that time can be represented by the real numbers (without any preference for their sign) – thus defining a linear order of physical states or global "Nows". Deterministically, the size of an ensemble (the number of microscopic states, or an appropriate measure if this number is infinite) does not change, while its coarse-grained size would, in general. This is why Boltzmann's $H$, which up to a factor may be assumed to represent "negentropy" for a diluted gas, may change in time (see Sect. 2). It will decrease even under deterministic equations of motion *in the direction of calculation* – provided the fixed concept of coarse graining was used to define the input ensemble. Further conditions, studied in ergodic theory, are necessary to exclude exceptional cases that are usually of measure zero. There is no *a priori* reason to calculate only in the conventional "forward" direction of time, but this is *empirically* the only direction in which statistical arguments lead to useful results – thus indicating a strong asymmetry of the facts. For applications to cosmology let me emphasize that the conclusion of a conservation of exact ensemble entropy under deterministic equations of motion includes the situation of a deterministic expansion (in particular cosmic inflation, which has often been claimed to *cause* a low entropy condition).

Although irreversible phenomena are mostly local, the thermodynamical arrow of time seems to have a universal direction. Its origin has, therefore, usually been discussed in a cosmological context. For example, one may assume a special cosmic initial condition at the big bang. Boltzmann, who believed in an eternal universe, argued instead that a giant chance fluctuation must have occurred in the distant past to form a low entropy state. The "future" would then be any time direction away from such a low entropy state. Boltzmann's proposal seems to imply, though, that it would then be far more probable if the *present* state of the universe – including all memories – had just formed in a chance fluctuation, since this state would still possess very low, but much higher entropy than the extremely improbable state in the distant past (see, however, Sect. 2). This idea has been discussed anew by the name "Boltzmann brains" in some recent, still speculative cosmologies. If, on the other hand, the low entropy is related to the special conditions at the big bang, the thermodynamical time arrow would have to change direction in an oscillating universe, while opposite arrows in causally connected parts of the universe seem to be excluded for dynamical reasons.[2]

The arguments based on deterministic dynamics do not apply without modification under stochastic laws of motion. However, a stochastic law by itself does not necessarily characterize a direction in time. If all states at some time $t_1$ have two possible successors at the later time $t_2$, say, and if this law is defined on all states, then each successor must on average also have two dynamically possible predecessors at time $t_1$. This kind of law defines a time-asymmetric indeterminism only when *applied* to a genuine subset of possible initial states, while not restricting the set of final states. This would be just another way of applying the "double standard" that has been duly criticized by Hugh Price.[3] The asymmetry is not a consequence of the stochastic law itself (see Sect. 3.4 of TD and the concept of "forks of indeterminism" mentioned therein). On the other hand, even deterministic laws may be asymmetric, but this would not offer a possibility to explain the increase of entropy, as demonstrated by the Lorentz force of an external magnetic field, or by the *CP*-violating terms in unitary quantum dynamics. Formal time-reversal symmetry violation is in these cases compensated for by another symmetry violation, which may be either physical, such as a *CP* transformation, or just formal, such as complex conjugation in the Schrödinger equation.

Our world is known to obey quantum theory, which is characterized by an indeterminism occurring in measurements and other "quantum events". There is absolutely no consensus among physicists about the interpretation and even the precise dynamical role of this "irreversible coming into being" of the observed facts, such as the click of a counter. Has it to be regarded as a specific part of the dynamical laws (as assumed in the form of von Neumann's "first intervention" or more explicitly in collapse theories), as representing events that (according to Pauli) occur outside the laws of nature, as a "normal" increase of information (as claimed in the Copenhagen interpretation), as determined by hidden variables that are not counted in conventional ensemble entropy (as in Bohm's theory), or as the consequence of indeministically splitting observers (as in Everett's interpretation)? Some quantum cosmologists refer to initial uncertainty relations or "quantum fluctuations" in order to justify the stochastic evolution of their quantum universe, although a global quantum state is never required to be "uncertain" (only classical variables would be).

In the pragmatic Copenhagen interpretation, this problem is essentially circumvented by denying any microscopic reality, while other above-mentioned proposals suggest novel laws or concepts, of which only some can be confirmed or ruled out in principle. Although these various interpretations have drastic consequences for the resulting model of the universe, they play surprisingly almost no role in cosmology. For example, the thermodynamical arrow might be the consequence of a time-asymmetric collapse mechanism if this were part of the laws. In the Copenhagen interpretation, there simply "is no quantum world" – hence no complete and consistent cosmology. Most cosmological models are therefore based on classical concepts, just allowing for some "quantum corrections", while indeterministic master equations are often derived from unitary equations of motion by using certain "approximations" in analogy to classical statistical physics. Such equations may then even *appear to explain* stochastic and irreversible quantum events, although they are implicitly using them.

Much philosophical work has also been invested into the pseudo-problem of distinguishing between a block universe and an evolving universe (a world of being versus a world of becoming). However, these apparently different pictures describe only different perspectives of the same object. One should be aware that a block universe picture is not restricted to a physical context. Historians have always applied it to the past, although they never had doubts that Cesar crossed the Rubicon according to his free decision. Similarly, we can use space-time diagrams to represent the *actual* motion even in the case of an indeterministic law, and hence to describe *potential* histories (individual members of an ensemble of possible histories). In particular, a block universe picture has nothing specifically to do with the theory of relativity (except that it is particularly convenient in the case of Lorentz invariance).

## 2. The arrow in classical physics

It is essential to keep in mind that time-symmetric laws are perfectly compatible with asymmetric solutions. Almost all solutions of the fundamental equations of motion are time-asymmetric, while recurrence times for isolated bounded systems would exceed the age of the universe by enormous factors. The symmetry of the laws of motion requires only that for every asymmetric solution that is realized in nature there must mathematically – not necessarily physically – exist precisely another, time-reversed one. In reality, very few systems can be considered as being isolated.[4] This means that the reversed solution would require an exact time reversal of its complete environment – an argument that must then be extended to the whole causally connected region of our universe. An extremely small "perturbation" (small change of the state at some time) would with overwhelming probability turn a deterministic solution with decreasing into one with increasing entropy (in both directions of time).[2]

Remarkable is only that there are whole classes of asymmetric solutions that are found in abundance, while members of the reversed class are rarely or never observed. As an example, consider the question of retarded versus advanced Maxwell fields of a given type of source. This asymmetry may be understood as a consequence of the presence of absorbers (including the early radiation era of our universe). Absorbers are based on the thermodynamical arrow of time, since they describe the transition to thermal equilibrium between radiation and matter. So they produce "retarded shadows", which, when forming a complete spatial boundary, give rise to local initial conditions of no incoming radiation at frequencies above the thermal spectrum (see Chap. 2 of TD). But why do all absorbers absorb in one and the same direction of time only?

The precise microscopic states of systems consisting of many interacting constituents can hardly ever be known even in a classical world. So it is common practice to use an incomplete (coarse-grained) description for them. For example, a gas may be described by the mean phase space distribution $\rho_\mu(\boldsymbol{p},\boldsymbol{q},t)$ of its molecules. Its evolution in the forward direction of time is then successfully described by Boltzmann's stochastic collision equation. This asymmetric success must be a consequence of properties of the thereby neglected *correlations* between molecules, since the increase of Boltzmann's entropy $S_B$,

(1) $$S_B := -Nk_B \overline{\ln \rho_\mu} = -Nk_B \int \rho_\mu \ln \rho_\mu d^3p d^3q \ ,$$

where $k_B$ is Boltzmann's constant and $N$ the particle number, can be deterministically understood as a dynamical transformation of information represented by the $\mu$–space distribution into information about correlations. Both kinds of "negentropy" are described by the $6N$-dimensional $\Gamma$-space distribution $\rho_\Gamma$, whereby the analogously defined ensemble entropy $S_\Gamma$ does *not* change under deterministic dynamics. While dynamical models readily confirm that correlations produced in a scattering process remain irrelevant for an extremely long time, one has to assume asymmetrically that only "retarded correlations" are present. This absence of advanced correlations is even "probable", while the low-entropy initial condition that leads to retarded correlations – such as a special initial $\mu$-space distribution $\rho_\mu$ – is *not*. The retarded correlations would be essential for reproducing $\rho_\mu$ backwards in time. Explaining these asymmetric correlations by referring to "causality" would be begging the question.

There are many appropriate ways to distinguish between macroscopic and microscopic ("irrelevant") degrees of freedom. They can all be formally described by some idempotent "Zwanzig" operator $P$ that acts on the $\Gamma$-space distributions $\rho = \rho_\Gamma$ (see Sect. 3.2 of TD),

(2) $\qquad \rho = P_{rel}\, \rho + P_{irrel}\, \rho\ , \qquad$ with $\quad P_{rel}^2 = P_{rel} \quad$ and $\quad P_{irrel} = 1 - P_{rel}\ ,$

where the macroscopically relevant part, $\rho_{rel} = P_{rel}\rho$, defines a generalized "coarse-grained" distribution. Various such projections have been shown to be useful in different physical situations. Macroscopic properties are characterized by a certain robustness or controllability, but they may vary with the physical situation. For example, correlations between molecules or ions are robust and relevant in solid bodies, while the corresponding lattice vibrations can then mostly be treated thermally. Although the exact dynamics requires a coupling between $\rho_{rel}$ and $\rho_{irrel}$, there often exists a probabilistic effective "master equation" for $\rho_{rel}$ that reflects the dynamical future irrelevance of $\rho_{irrel}$ for the dynamics of $\rho_{rel}$, as exemplified by Boltzmann's collision equation, where $\rho_{rel}$ can be defined in terms of $\rho_\mu$.

The physically appropriate relevance concept used to define $\rho_{rel}$ may thus change in time. In such cases, the usual ignorance of microscopic degrees of freedom can be deterministically transformed into "lacking information" about arising macroscopic ("relevant") ones – such as the positions of droplets formed during a condensation process. This happens, in particular, in symmetry-breaking phase transitions, or in measurements of microscopic variables (but these processes assume a completely new form in quantum theory). Strictly speaking, only the complete ensemble entropy, measured by the mean logarithm of $\rho_\Gamma$ itself (without any coarse-graining), is conserved under deterministic equations of motion. Physical entropy is usually defined *not* to include that part which represents lacking information about, but rather as a *function of given* macroscopic variables. However, the transformation of physical entropy into entropy of lacking information cannot be used in a cyclic process to construct a perpetuum mobile of the second kind.[5] The formal entropy of lacking information is in general thermodynamically negligible, but it may become essential for fundamental considerations – such as those involving Maxwell's demon. While the precise definition of entropy (its specific relevance concept or Zwanzig projection) is in principle a matter of convenience, the initial cosmic low-entropy condition that would "cause" an arrow of time must represent a real property of the universe, and its precise nature should therefore be revealed.

The robustness of macroscopic properties together with the retardation of all correlations between them means that there are many redundant "documents" (including fossils and personal memories) about the macroscopic past. The macroscopic past is said to be "overdetermined" by the macroscopic present or future.[6] It appears fixed because it could not have been different if just one or a few documents were found to be different. Precisely this consistency of the documents makes them trustworthy and distinguishes them from mere chance fluctuations with the same low value of physical entropy. Julian Barbour has called states that contain consistent documents (regardless of their causal origin) "time capsules".[7] Since conventional concepts of physical entropy are local (based on an entropy density), they cannot distinguish between consistent and inconsistent documents. An evolved ("historical") state has much lower statistical probability than indicated by its physical entropy, and so may rule out Boltzmann brains for being "unreasonable" (see Sect. 3.5 of TD).

In most cosmological models, the low-entropy initial condition is represented by a "simple" state of high symmetry – very different from a later state of low but larger entropy that describes complexity and dynamical order (as it exists in organisms, for example). While an exactly symmetric state could not evolve into an asymmetric one by means of symmetric and deterministic laws, a state consisting of classical particles cannot be exactly (microscopically) homogenous. The information capacity of a single continuous variable is already infinite, and its exact value would violate this symmetry. Nonetheless, a Laplacean model universe with small initial asymmetries that determine all later arising complexity may be consistent, but not very convincing in an evolutionary picture.

While, in a laboratory situation, thermal equilibrium normally requires macroscopically homogeneous ensembles, such states are extremely improbable (and hence unstable) in self-gravitating systems. Gravitating stars and galaxies, for example, possess negative heat capacity: they become hotter when losing energy (see Sect. 5.1 of TD). Classically, this negative heat capacity would even be unbounded. Therefore, the initial homogeneity of the universe becomes a major candidate for the specific low entropy condition that characterizes this universe. Roger Penrose has formulated this condition in general relativity by postulating a vanishing Weyl tensor on all past singularities. This source-free part of the spacetime curvature tensor can be interpreted as representing gravitational radiation. The Weyl condition would thus mean that all gravitational radiation must be retarded (possess sources in its cosmic past). An analogous condition had been proposed for electromagnetic radiation by Planck in a debate with Boltzmann, and later by Ritz in a debate with Einstein. However, because of the weak coupling of gravity to matter, the Weyl tensor condition cannot similarly be explained by the thermodynamic properties of absorbing matter. It may then itself have to establish the causal nature of the universe, that is, it would have to be responsible for the absence of future-relevant initial correlations.

**3. The arrow in quantum theory and quantum cosmology**

Although the quantum formalism of irreversible processes is formally quite analogous to its classical counterpart (see Sect. 4.1 of TD), there are at least three genuine quantum aspects that are important for the arrow of time: (1) the superposition principle, (2) a quantum indeterminism of controversial origin – often described by a collapse of the wave function, and (3) quantum nonlocality – a specific consequence of (1).

The superposition principle allows *exactly* symmetric states for all kinds of symmetries. Such symmetric states may then form candidates for an entirely unspecific initial pure state. Although they cannot unitarily evolve into asymmetric states by means of a symmetric Hamiltonian, they could do so by means of an appropriate indeterministic collapse of the wave function that is in conflict with the principle of sufficient reason. While such a collapse has always to be used *in practice* to describe measurements or phase transitions in terms of quantum states, a non-unitary modification of the Schrödinger equation that would satisfactorily describe it has never been experimentally confirmed. Therefore, Everett's "branching" of the quantum universe (including all observers) into different autonomous components describing quasi-classical "worlds" must be taken seriously as an alternative. The branching can be specified by means of an in practice irreversible process of decoherence that is in accordance with the Schrödinger equation.

While the collapse would define a time-asymmetric law, the time arrow of decoherence (retarded entanglement) must again be explained by means of an initial condition – now for the global wave function. A universally valid Schrödinger equation does in principle also admit the anti-causal process of recoherence, but this would be very rare under an appropriate initial condition. Although any initial symmetry of the global state must be conserved under a symmetric Hamiltonian, an unentangled ("simple") symmetric state can evolve into a sym-metric superposition of many asymmetric Everett branches possessing a complex structure (independent "worlds"). This subtlety is neglected in most cosmological models – in particular when other Everett branches are disregarded for being "meaningless". Calculating backwards in time, however, would require knowledge of *all* Everett branches or collapse components (including the unobserved ones) and their phase relations as an input.

A stochastic collapse by itself (that is, when neglecting the accompanying decoherence processes) would *reduce* nonlocal entanglement, since it is usually defined to select components that factorize in the relevant subsystems (see Sects. 4.6 and 6.1 of TD). This consequence applies as well to the transition into an *individual* Everett world that is experienced by local (themselves branching) observers which are in definite states. Such a reduction of entanglement is required, in particular, in order to obtain definite outcomes in measurement-like processes, or to allow the preparation of pure initial states in the laboratory or during a process of self-organization.

This indeterministic transition into less entangled states must reduce any physical entropy measure that is defined by means of a Zwanzig projection of locality (required if entropy is to be an extensive quantity). It is here important to recognize the difference between *classical* microscopic states, which are local by definition (that is, they are defined by the states of all their local subsystems), and generically nonlocal quantum states. Therefore, the physical (local) entropy of a completely defined ("real") classical state is trivial (minus infinity), while that of a pure quantum state is not only non-negative, but in general also much greater than zero (non-trivial). The permanent creation of uncontrollable quantum entanglement by decoherence is the dominating source of physical entropy production, which finally leads to apparent ensembles (improper mixtures) that represent thermodynamical equilibrium. It is tacitly used in "open systems" quantum mechanics. The *reduction* of entropy in a process of symmetry breaking, on the other hand, is usually very small in comparison with thermodynamic entropy, but it may be cosmologically essential when, for example, it leads to new Goldstone type particles with their enormous entropy capacity (see Sect. 6.1 of TD).

Another novel consequence of quantum theory that regards the arrow of time is, of course, the entropy bound that governs gravitational contraction. It is characterized by the Bekenstein-Hawking black hole entropy, given by

(3) $$S_{BH} = 4\pi \frac{k_B G M^2}{\hbar c}$$

for spherical and electrically neutral black holes. Here, *G* is the gravitational constant. The fact that $S_{BH}$ is quadratic in the mass *M* indicates that it describes some kind of correlations. According to classical general relativity, spacetime geometry is regular at the black hole horizon, while there has to be a future singularity inside. However, the interior cannot causally affect the external region: it must for all times remain in our future. This leaves much freedom for the unknowable physics inside. In particular, *quantum* gravity does not allow distinguishing between past and future singularities any more (see below). Therefore, one can only postulate a Weyl tensor condition on *all* space-like singularities. Such a time-symmetric condition is not only compatible with all observations – it may even prevent black hole interiors and horizons to form (thus avoiding any genuine information loss paradox).[8]

Most of these genuine quantum aspects of the cosmic arrow of time have so far received little attention – perhaps because they depend on the interpretation of the quantum formalism. Cosmological models are mostly presented in classical terms rather than in terms of quantum states (their superpositions). In particular, arguments using Feynman's path integral often replace this integral, which describes a superposition of paths (precisely equivalent to a wave function[9]), by an *ensemble* of paths in classical configuration space. Selecting a sub-ensemble or an individual path from them is nonetheless equivalent to a time-asymmetric collapse of the wave function. Similar objections apply to using tunneling probabilities, since any decay process must quantum mechanically be described as a coherent superposition of different decay times as long as the corresponding partial waves are not irreversibly decohered from one another – thereby letting the decay event appear to become "real" (Sect. 4.5 of TD).

A consistent quantum description requires that classical general relativity is replaced by quantum gravity. This does not necessarily require a complete understanding of this theory. Although the problem of the arrow of time would certainly remain essential in an ultimate theory, the precise meaning and validity of existing proposals (such as string theories) remain highly speculative as yet. Standard quantization of the canonical form[10] of General Relativity in the Schrödinger picture, on the other hand, leads to the Wheeler-DeWitt equation (or Hamiltonian quantum constraint),[11]

(4) $\quad\quad\quad\quad\quad\quad H \Psi = 0$ ,

which may be expected to form an effective theory of quantum gravity at "low" (that is, normal) energies. The wave functional $\Psi$ depends on spatial geometries and matter fields on arbitrary simultaneities. Since the Schrödinger equation now takes the form $\partial_t \Psi = 0$, there exists no time parameter any more that could be used to formulate a direction in time. This "timelessness" has occasionally been regarded as a severe blow to this approach, although it must apply to all quantum theories that are reparametrization invariant in their classical form. However, the physical concept of time – and even its arrow – can be recovered in a satisfactory way under very reasonable assumptions.[12]

The first important observation for this purpose is that the Wheeler-DeWitt equation for Friedmann type universes is globally of hyperbolic type, with a time-like variable $\alpha := \ln a$, where $a$ is the cosmic expansion parameter.[13] This fact defines an intrinsic "initial" value problem in $\alpha$ or $a$, for example at the big bang ($\alpha = -\infty$), which is now identical with a big crunch. The Wheeler-DeWitt equation is drastically asymmetric under a change of sign of $\alpha$, thus suggesting an asymmetric solution without explicitly postulating it by means of asymmetric boundary conditions. The second step for recovering conventional time is a Born-Oppenheimer expansion in terms of the inverse Planck mass, which equals $1.3 \cdot 10^{19}$ proton masses.[14] This mass characterizes all geometric degrees of freedom. The expansion leads to an approximately autonomous evolution of partial Wheeler-DeWitt wave functions for the matter degrees of freedom along WKB trajectories that are defined in most regions of the configuration space of geometries. This is analogous to the adiabatic evolution of electron wave functions along classical orbits of the heavy nuclei in large molecules. This evolution has precisely the form of a time-dependent Schrödinger equation (plus very small corrections).[15] The in this way recovered concept of time represents arbitrary time coordinates for all possible foliations, independently for all dynamically arising quasi-classical spacetimes (branches).

Note that this WKB approximation does not by itself justify an *ensemble* of trajectories, since it preserves the global superposition of all partial waves that evolve along different WKB trajectories – just as small molecules (for which the positions of nuclei are *not* decohered to become quasi-classical variables) are known to exist in energy eigenstates in spite of the validity of the WKB approximation for them. However, since observers have different states in these different autonomous partial waves, they can only observe their own "branch" as an evolving quantum world. The global intrinsic dynamics would be required, though, to *derive* initial conditions for all partial Schrödinger waves, to be used in the WKB region of geometry (at some distance from the big bang).

According to loop quantum cosmology, the Wheeler-DeWitt equation (in this theory replaced by a difference equation with respect to $a$) can be continued through $a = 0$ to negative values of $a$.[16] The configuration space of three-geometries is in this way duplicated by letting the volume measure assume negative values (turning space "inside out" while going through $a = 0$). Since the Hamiltonian does not depend on the newly invented sign of $a$, however, the Wheeler-DeWitt wave function must be expected to be symmetric under this parity transformation, too. Its continuation would then have to be interpreted as an added superposition of other physically *expanding* universes. Since the WKB times, which represent classical time, can *not* be continued through $a = 0$, the interpretation of negative values of $a$ as representing pre-big-bang times is highly questionable. The fundamental arrow, including its consequence of decoherence outside the validity of a WKB approximation, must depend on some low entropy "initial" condition in $a$ for all other ("spacelike") degrees of freedom that occur as physical arguments of the Wheeler-DeWitt wave function. It would be hard to understand how the low entropy state at $a = 0$ could have been "preceded" by an even lower entropy at $a < 0$ in order to avoid a reversal of the thermodynamical arrow in the classical picture of an oscillating universe.

In spite of the success in recovering physical time for the autonomous Everett branches that represent quasi-classical spacetimes, "timelessness" has recently become a hot issue based on severe misunderstandings. It has even been used as a motivation to present obscure and speculative solutions to this non-existing problem. I will, therefore, now give a brief review of *different* concepts of timelessness that have been used in this connection.

**4. A brief history of timelessness**

Newton described planetary motions in the time-dependent form $r(t)$ and $\varphi(t)$. He also assumed, by means of his laws, that absolute time $t$ can be read from appropriate clocks, such as the rotation of the Earth, $\alpha = \omega t$. Elimination of $t$ from the first two functions leads to Kepler's orbits $r(\varphi)$, or that from all three functions to a clock dependence $r(\alpha)$ and $\varphi(\alpha)$. This trivial elimination of time has recently been used by some authors to argue that one should "forget time" in all dynamical considerations.[17] However, this argument completely neglects the fact that it is precisely Newton's time that simplifies his laws of motion, as has been clearly emphasized already by Henri Poincaré. So, in Newtonian physics there is a preferred time parameter that could indeed be interpreted as representing "absolute" time.

The concept of absolute time was not only questioned for philosophical reasons by Leibniz and Mach, it also lost its empirical justification in General Relativity. In Special Relativity, absolute time is replaced by an absolute spacetime metric that still defines path-dependent proper times. According to the principle of relativity, they control all physical motions in the same preferred way as Newtonian time did in non-relativistic physics. In particular, local clocks measure proper times along their world lines, but the spacetime metric can be assumed to exist even in the absence of physical clocks.

In General Relativity, the spatial metric defined on arbitrary simultaneities becomes itself a dynamical object[10] – just as any matter field. Its evolution gives rise to a succession of spatial curvatures that defines a foliation of spacetime. It can be parametrized by an arbitrarily chosen time coordinate, but there is no *preferred* coordinate or time parameter any more. Julian Barbour has discussed this Machian property, which he called timelessness, in great detail, including many consequences that were historically important.[18] However, the arising metric still defines proper times for all world lines (Wheeler's *many-fingered time*), and the evolving spatial metric can itself be regarded as a many-fingered *physical* clock.[19] Although there are many different time-like foliations of the same spacetime, each one defines a parametrizable succession of states, and this dynamical construction allows in general the formulation of a unique initial value problem – hence an initial condition of low entropy. There are also mathematically consistent non-relativistic Machian ("relational") theories.[20]

The complete absence of any time parameter from the Wheeler-DeWitt wave function (genuine timelessness), discussed in the previous section, is a specific quantum property: it is a consequence of the fact that in quantum theory there are no trajectories (in configuration space) that could be parametrized. Hence, in quantum gravity there are no classical space-times that could give rise to a time-like foliation. There is only a probability amplitude for spatial geometries (many-fingered physical clocks) entangled with matter fields.[21] For the same reason, the concept of "relational observables" is inappropriate, since it uses a classical picture of orbits, required to define such relations between variables.[17] The entanglement describes also the decoherence of macroscopically different geometries from one another if matter is regarded as an environment to geometry.[12] Among the parameters characterizing these spatial geometries is the "intrinsic time" $\alpha = \ln a$. It is remarkable that this genuine timelessness (the inapplicability of *any* external time parameter) was known before weaker non-quantum forms were discussed by this ambitious name. It seems to have initially been mostly regarded as a formal problem. The reason may be that early physicists working on quantum gravity did not take the Wheeler-DeWitt wave function seriously as representing reality. They either used semi-classical approximations for its interpretation (even where they did not apply), or they preferred a Heisenberg picture, in which the problem is less obvious.[22]